\documentclass[prl,amsmath,aps,twocolumn,superscriptaddress]{revtex4-1}
 \usepackage{graphicx,xspace}
 \usepackage{float}
 \usepackage{amsmath}
 \usepackage{amssymb}
 \usepackage{color}
 \definecolor{c1}{rgb}{1, 0, 0}
 \definecolor{c2}{rgb}{0, 1, 0}
 \definecolor{c3}{rgb}{0, 0, 1}
 \definecolor{c4}{rgb}{1, 0, 1}
 \definecolor{c5}{rgb}{0, 1, 1}
 \usepackage[normalem]{ulem}
 \input{epsf}
 \begin{document}
 \newcommand{\scalar}[2]{\left \langle#1\ #2\right \rangle}
 \newcommand{\me}{\mathrm{e}}
 \newcommand{\mi}{\mathrm{i}}
 \newcommand{\dif}{\mathrm{d}}
 \newcommand{\period}{\text{per}}
 \newcommand{\free}{\text{fr}}
 \newcommand{\mq}[2]{\uwave{#1}\marginpar{#2}} 
 \include{latexcommands}
 
 \title{Critical interfaces and duality in the Ashkin Teller model}
 \author{Marco Picco} \email{marco.picco@lpthe.jussieu.fr}
 \affiliation{CNRS, LPTHE, Universit\'e Pierre et Marie Curie, UMR 7589, 4 place
Jussieu, 75252 Paris cedex 05, France} 
 \author{Raoul Santachiara} \email{raoul.santachiara@lptms.u-psud.fr}
 \affiliation{CNRS, LPTMS,
              Universit\'e Paris-Sud, UMR 8626, B\^atiment 100, 91405 Orsay, France}
   
   \begin{abstract}
     We report on the numerical measures on different spin interfaces
     and FK cluster boundaries in the Askhin-Teller (AT) model.  For a
     general point on the AT critical line, we find that the fractal
     dimension of a generic spin cluster interface can take one of four
     different possible values. In particular we found spin interfaces
     whose fractal dimension is $d_f=3/2$ all along the critical
     line. Further, the fractal dimension of the boundaries of FK
     clusters were found to satisfy all along the AT critical line a
     duality relation with the fractal dimension of their outer
     boundaries. This result provides a clear numerical evidence that
     such duality, which is well known in the case of the $O(n)$ model,
     exists in a extended CFT.
   \end{abstract}
 
 \maketitle
 \underline{Introduction}
 
 These last years have seen the study of geometrical objects in
 two-dimensional critical statistical models as one of the most active
 areas in statistical and mathematical physics.  The universality
 classes of a large variety of critical models (critical percolation,
 the self-avoiding walks, loop erased random walk, $q$-states Potts
 models) can be described in terms of one parameter family of loop
 models, the $O(n)$ loop models.
clusters which appear in the  
 Combining methods of conformal field theory (CFT), with a Coulomb-gas
 representation \cite{Nienhuis_CG}, all geometrical exponents
 characterizing the fractal geometry of the $O(n)$ critical loops can
 be computed \cite{SalDupl}.  The interest in these investigations has
 been certainly boosted by the more recent discovery that the continuum
 limit of certain boundary loops (i.e. loops created by imposing
 certain boundary conditions) can be described by conformally invariant
 stochastic growth, the so called Schramm-Loewner evolutions (SLE). The
 SLE approach, besides providing new formulas, paved also the way to
 put the results obtained from CFT methods on a more firm mathematical
 foundation.
 
 According to these results, which concern mainly the $O(n)$ models in their
 critical dense phase, it is tantalizing to suggest a 
 completely new interpretation of
 critical phases in terms of concepts of stochastic geometry.  Still,
 this scenario is far to be established.  The critical point of 2D
 systems can be classified according to the different families of
 CFTs. In this respect, the $O(n)$ model are described by the most
 simple of CFT family, i.e. the one constructed from the conformal
 symmetry alone. Other lattice statistical models, typically
 characterized by some symmetry in the internal degree of freedom, have
 critical points described by the so-called extended CFT, i.e. CFTs
 with additional symmetries.  A representative example are the $Z_N$
 spin lattice models where the spins take $N$ values and interact via a
 nearest-neighbor potential which is invariant under a $Z_N$ cyclic
 permutation of the $N$ states.  These spin models have critical points
 which, for $N\geq 4$, are described by a family of extended CFT, the
 so-called parafermionic CFT \cite{Zamo1}. If the role of the $Z_N$ additional
 symmetries is under control as far as the operator algebra or the
 classification of the conformal boundary conditions is concerned, its
 effects on the geometrical features of the corresponding critical
 phases is still not understood.  The study of the geometrical features
 of the extended CFTs turns out to be a very hard problem. Some
 progress in this direction has been done by defining loop models
 associated to some extended CFTs \cite{Fendley} or by proposing a possible extension
 of the SLE approach to these CFTs\cite{Rasmu,Ch}. But, so far, the most important
 insights into this problem come from numerical measurements of the
 fractal dimensions associated to the spin interfaces for $Z_4$ and
 $Z_5$ spin models \cite{PS}. By measuring systematically all the
 different bulk and boundary spin interfaces, it was found that there
 is a limited number of possible values for the fractal dimension which
 can be understood on the basis of the classification of $Z_N$
 conformal boundary conditions \cite{PS2}. One of these value,
 corresponding to a certain interface, was found in agreement with the
 one proposed on the basis of CFT computation in \cite{R}.  This
 scenario, which establishes for the first time a connection between
 geometrical objects and extended CFT, has been made particularly clear
 in \cite{PS2} where the spin cluster interfaces of the $Z_4$ model
 model were studied.
 
 The critical point of the $Z_4$ is particularly interesting as it
 represents a particular point on the self-dual critical line of the
 well known Ashkin-Teller (AT) model \cite{AT}. The phase diagram of
 the AT model, defined below, presents a critical line which is
 described by the compactified free Gaussian boson along which the
 critical exponents change with the compactification radius.
 
 In this paper we report on the numerical measures of the fractal
 dimension of different spin interfaces of the AT model by showing how
 the results found in \cite{PS2} generalise all along the critical
 line.  In particular we found a spin interface whose fractal dimension
 is $d_f=3/2$ all along the critical line suggesting this interface to
 be described by an SLE$_4$ process. Moreover, we measured the fractal
 dimension of particular FK cluster defined bellow. Interestingly,
 these values seem to satisfy all along the critical line a duality
 relation with the fractal dimension of one particular spin interface.
 This is a clear numerical evidence that such duality, which is well
 known in the case of the $O(n)$ model, exists in a extended CFT.  Note
 that the same duality was also found by studying spins and FK
 interfaces in the Potts model at the random conformal critical
 point \cite{Pottsdis} which are also believed to be described by
 (non-unitary) extended CFTs. In the case of $Z_4$ spin model, a
 duality was predicted on the basis of a CFT computation but the
 proposed FK cluster was not the same as the one studied here.
 
 \underline{The model}

 The Ashkin-Teller model is usually defined in terms of two coupled
 Ising models. On each site $i$ of a square lattice one associates a
 pair of spins, denoted by $\sigma_i$ and $\tau_i$, which take two
 values, say up (+) and down (-). The Hamiltonian is defined by
 \begin{equation}
 H_{AT}(\{\sigma_i,\tau_i\}) =-\sum_{<ij>} \left(  K (  \sigma_i \sigma_j + \tau_i
\tau_j )  
 + K_4 \sigma_i \sigma_j \tau_i \tau_j \right) \; .
 \label{HAT}
 \end{equation}
 The AT model presents a rich phase diagram which has been very
 well studied \cite{baxter,N}: there is a critical line defined by the
 self-dual condition $\sinh 2 K = \exp (-2 K_4)$ and terminating at
 $K=K_4=K^{P}$ where $K^{P}=\ln 3/4$. The point $K^{P}$ corresponds to
 the $4$-states Potts model critical point.  Other two special points
 are known: i) the point
 ($K_4=K_4^{I}=0$,$K=K^{I}=\ln\left(1+\sqrt{2}\right)/2$) corresponding
 to two decoupled critical Ising models and ii) the so called Fateev
 Zamolodchikov (FZ) point, located between the $K^{P}$ and $K^{I}$, and
 described in the continuum limit by the $Z_4$ parafermionic theory
 mentioned above.
 
 For our purposes, it is convenient to rewrite the model (\ref{HAT}) in terms of the  
 spin variables $S_i$ which take four values, 
 $S_i=1,2,3,4$ and defined via the correspondence:
 \begin{eqnarray} 
   S_i=1 :  (\sigma_i,\tau_i) = (+,+) &;&  S_i=2 : (\sigma_i,\tau_i) = (+,-) \nonumber \\ 
 S_i=3 : (\sigma_i,\tau_i) = (-,-) &;&  S_i=4 : (\sigma_i,\tau_i) = (-,+) \; 
 \label{HATS} 
 \end{eqnarray}
 With this formulation, it is easy to see that the $Z_4$ symmetry of
 the model (\ref{HAT}) is completely explicit.
 
 \underline{Spin cluster interfaces}
  
 In this paper we investigated numerically the spin cluster interfaces
 of the AT model which are defined, in the representation (\ref{HATS}),
 as boundaries of cluster of spins $S_i$.  A spin cluster indicates a
 connected set of spins which can take a given set of values.
 
\begin{figure}[h]
 \includegraphics[scale=1.0,angle=0]{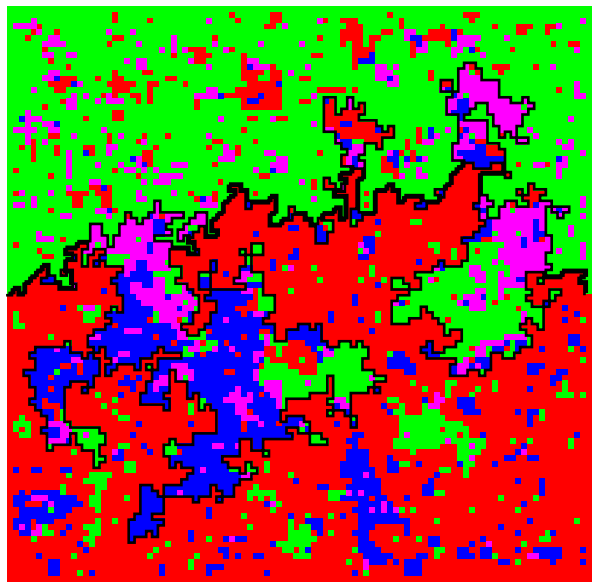}
 \caption{Spin interfaces for one sample of size $100 \times 100 $ with the
boundary condition $(1|2)$.
 {\color{c1} Red} color corresponds to $S=1$; {\color{c2} Green} color corresponds
to $S=2$;
 {\color{c3} Blue} color corresponds to $S=3$; {\color{c4} Magenta} color
corresponds to $S=4$
 }
 \label{Plot100}
 \end{figure}

 In analogy with the chordal SLE interfaces defined in the critical
 $O(n)$ model, we consider the model on a bounded domain and we define
 spin interfaces generated by imposing certain spin configurations on
 the boundary.  More in detail, we simulate finite square lattices of
 size $L \times L$ with certain boundary conditions $(A_1 + A_2
 \cdots|B_1 + B_2 \cdots)$. By this notation, we mean that we set one
 half of the boundary spins to take the values $A_1,A_2,..$ and the
 other half the values $B_1,B_2..$ with equal probabilities. Moreover,
 we impose that the change from one condition to the other is on the
 middle of the two opposite borders of the square lattice. Then, for
 each spin configuration, there is at least one interface defined as
 the line on the dual lattice separating the $A_i$ spins connected to
 one boundary from the $B_i$ spins connected to the other boundary. We
 present one example in Fig.~\ref{Plot100} which correspond to the
 boundary condition $(1|2)$. In this figure there is two interfaces,
 one separating the {\color{c1} red} spins corresponding to $S=1$ and
 connected to one boundary from other colors, the second interface
 separating the {\color{c2} green} spins corresponding to $S=2$ and
 connected to another boundary from other colors. These interfaces are
 shown as thin black line.  The common part of these interfaces is
 shown as a thick black line.
 
 One can easily observe that, using the $S_i$ spin degree of freedom,
 the $Z_4$ symmetry is explicit in the definition of the interfaces and
 is crucial to understand the properties of these interfaces. In this
 respect, the FZ point plays a special role: this is the only point of
 the AT critical line where, in the continuum limit, the $Z_4$ symmetry
 conserved currents form a chiral algebra. The correspondent CFT (i.e.
 the $Z_4$ parafermionic CFT) enjoys the properties i) to have a finite
 number of primary operators which close under operator algebra and ii)
 the Hilbert space can be classified on the basis of the $Z_4$
 transformation properties of the states.  Using these properties we
 were able to almost fully characterize the conformal boundary
 conditions in terms of spin configurations. Despite the fact we
 considered a great number of different boundary spin interfaces, we
 found a limited number of fractal dimensions and this could be
 understood on the basis of such classification. These results were
 confirmed by simulating also for bulk spin interfaces i.e. closed
 interfaces surrounding spin clusters as opposed to open interfaces
 connecting opposite boundaries.
  
\begin{figure}[h]
 \includegraphics[scale=1.0,angle=0]{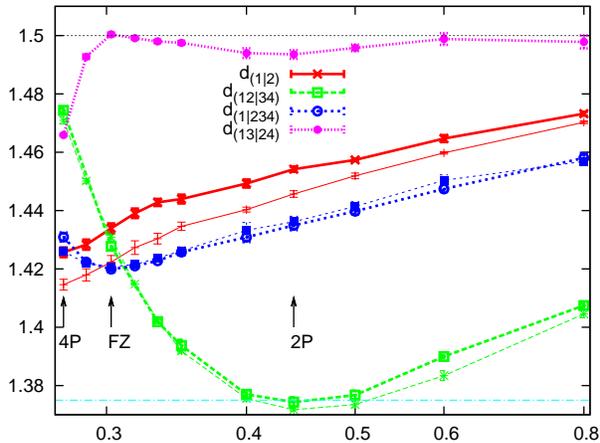}
 \caption{Fractal dimensions for spin interfaces vs. $K$. We show the
   fractal dimensions for four types of interface with two sizes
   ranges. A thin plot is used for a fit with sizes $L=20-320$. For
   $(1|2)$ and $(12|34)$, the thick plot corresponds to the fit with
   sizes $L=80-1280$. For $(1|234)$ the thick plot is for
   $L=40-640$. For $(13|24)$ we were able to simulate only up to the
   size $L=320$. The critical points corresponding to the $4$-states Potts (4P), the FZ and to the two decoupled Ising (2P) models are indicated.}
 \label{Plot2AT}  
 \end{figure}

 We present here how the numerical results obtained for the FZ point
 generalize to the entire critical line. The first main observation is
 that, for a general point on this line, the fractal dimension of many
 different boundary interfaces take only one of four different values.
 In Fig.~\ref{Plot2AT} we show the values of the fractal dimensions of
 four representative spin interfaces computed for different points on
 the critical line.  The interface $(12|34)$ has been already
 considered in \cite{CLR}. In the $\sigma$, $\tau$ variables this
 conditions is equivalent to impose $\sigma=+$ ($\sigma=-$) on the left
 (right) border of the lattice while $\tau=+, -$ is free. At the Ising
 decoupling point, the $(12|34)$ interface is thus equivalent to the
 $SLE_{3}$ Ising interface. Its fractal dimension $d_{(12|34)}$ has to
 be equal to $d_{(12|34)}=1+3/8$, in perfect agreement with the value
 shown in Fig.~\ref{Plot2AT}.  Note that at the Ising decoupling point,
 we found three other non-trivial fractal dimensions, $d_{(1|234)}$,
 $d_{(1|2)}$ and $d_{(13|24)}$. From the classification done in \cite{AO}, one can
show that the  
 boundary operator generating  the 
 $(13|24)$ conditions has conformal dimension $1/4$. The corresponding interface is
thus expected \cite{Cardy_c1,GW,HBB} 
 to be described by an $SLE_4$  process and therefore to have fractal dimension
dimension $3/2$,
  again in agreement  with the value shown in Fig.~\ref{Plot2AT}. 
 On the contrary,  we do not have any theoretical argument to explain  
 the other two values $d_{(1|234)}$,
 $d_{(1|2)}$ at the Ising decoupling point. These boundary
 conditions are indeed highly non trivial in terms of the Ising $\sigma$ and
 $\tau$ variables.
  
 At the FZ point, on the basis of our boundary classification, we
 conjectured \cite{PS2} that the value $d_{(12|34)}$ should be equal to
 the one of the $(1|234)$ interfaces which was predicted to be
 $d_{(1|234)}=17/12$ in \cite{R}, in agreement with the numerical
 findings \cite{PS2}. From Fig.~\ref{Plot2AT} one can notice that these
 two fractal dimension are slightly different, but this difference can
 be explained by finite size effects, as shown in
 Fig.~\ref{Plot2AT}. In this figure, we represent by thin plots the
 fractal dimensions obtained by fiting the numerical data for sizes
 $L=20-320$. This can be compared with the data obtained for sizes
 $L=80-1280$ for the $(1|2)$ and $(12|34)$ boundary condition cases and
 represented as thick lines. For the boundary condition $(1|234)$, we
 were able to simulate only up to size $L=640$, the thick line for this
 case corresponds then to a fit in the range $L=40-640$. For the
 boundary condition $(13|24)$, we were able to simulate only up to size
 $L=320$.
 
 The finite size effects become more and more important when
 approaching the four states Potts model. At this point, the model has
 a permutational $S_4$ symmetry (and thus larger than a $Z_4$ one) and
 the Boltzmann weight associated to an interface can only take two
 values depending if the spin at the interface are equal or not. It is
 well known that the critical point of a $q\leq 4$-states Potts model
 is in the same universality class as the $O(\sqrt{q})$ model in its
 dense phase. The $4$-states Potts model in particular is described by
 the point separating the dense and dilute critical phases of the
 $O(2)$ loop model. The spin boundary interfaces and the
 Fortuin-Kasteleyn (FK) cluster interfaces (discussed below) of the
 $4$-states Potts models are predicted to have the same fractal
 dimension $d_{f}=3/2$. At the Potts model we expect the fractal
 dimension of all the spin interfaces to take this value. But it is
 well known that strong logarithmic corrections exist for the $4$-states
 Potts model \cite{CNS} and thus it is not a surprise that we obtain numerical
 values not in very good agreement with this prediction. In a direct
 measurement of interfaces of spin clusters for the $4$-states Potts
 model in \cite{ZS}, a fractal dimension in very good agreement with
 our value $d_{(1|234)}$ was obtained.
 
 The interface $(13|24)$ is particularly interesting: from our
 numerical analysis, a fractal dimension with the value
 $d_{(13|24)}=3/2$ is obtained all along the critical line (apart close
 to the limit of the $4$-states Potts model where strong finite size
 effect are expected as explained above). To support this result, we have verified,
by a  
 transfer matrix numerical analysis of the  boundary states 
 that the conformal dimension of the operator associated to the conditions 
 $(13|24)$ is compatible with the  dimension $1/4$ all along the critical line.
Moreover, 
 one can easily show that the corresponding bulk interfaces can be defined as the
boundaries of the
 spin clusters which appear in the high-temperature expansion of the AT
 model \cite{Nienhuis_CG,N,Saleur_AT}.  Very recently,  theoretical arguments to
prove that 
 these interfaces have dimension $3/2$ 
  have been proposed in
 \cite{YM}.
 
 \underline{FK interfaces} Although the spin interfaces are a very
 natural object, it is often very difficult if not impossible to tackle
 the study of these interfaces by some exact theoretical method. This
 is the case of the $q=3,4$ Potts model where the geometric features of
 spin interfaces boundaries are by far less understood than the FK
 cluster boundaries \cite{DJS}. The FK cluster are at the basis of
 Fortuin-Kasteleyn and their boundaries are directly related to the
 critical loops of the $O(n=\sqrt{q})$ model.  The FK cluster are
 generally obtained by a random walk on a spin cluster and therefore
 the relation between these two kinds of clusters is highly nontrivial.
 The main insight into the properties of spin interfaces comes from the
 observation that the critical exponents extracted by studying a spin
 cluster in a $q$ Potts model characterize the universality class of
 the FK cluster of correspondent tri-critical Potts model
 \cite{SV,V}. According to this, the fractal dimension of the spin
 cluster boundary $d_{s}$ and of the associated FK cluster $d_{FK}$
 boundary are related by the Duplantier duality relation:
 \begin{equation}
 (1-d_s)(1-d_{FK})=\frac{1}{4} \; .
 \label{Duality}
 \end{equation}
 In the Coulomb gas formulation of the Potts model, this duality can be
 expressed in terms of an electric-magnetic duality
 transformation. This transformation also relates the descriptions of
 the dilute and dense phase of the correspondent $O(n)$ model \cite{GW}.
 \begin{figure}[h]
 \includegraphics[scale=1.0,angle=0]{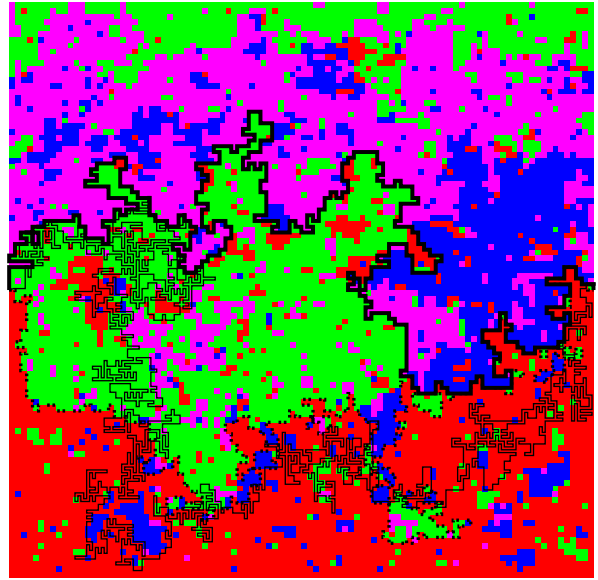}
 \caption{One sample of size $100 \times 100 $ with the boundary
   condition $(1|234)$.  The different colors correspond to different
   values of spin as in Fig.~\ref{Plot100}. The thin interface
   corresponds to the FK interface constructed as described in the
   text, the dotted interface corresponds to the spin interface
   $(1|234)$ and the thick interface corresponds to the outer
   boundary. Note that the outer boundary of the
   FK cluster separates the spin $S=1,2$ from the spins $S=3,4$: it's
   fractal dimension is then given by $d_{(12|34)}$} 
 \label{FKvsoth}  
 \end{figure}

 At our knowledge, the first numerical evidence that a duality relation
 relating FK cluster to spin cluster exists for extended CFTs has been
 presented in \cite{Pottsdis} by studying a $3$-states Potts model at the critical
 random fixed point.  A theoretical argument that Duplantier duality
 still holds for extended CFTs has been proposed in \cite{R} on the
 basis of CFT results.  In \cite{PSS} we attempted to define
 generalised FK clusters for the $Z_4$ and $Z_5$
 spin models: we observed that these clusters do not percolate at the
 critical point of the model under consideration.
 
 There exist in fact another type of FK clusters for the AT
 model. These clusters are constructed by considering one of the two
 coupled Ising models appearing in (\ref{HAT}). By rewriting the
 Hamiltonian (\ref{HAT}) in the following way:
 \begin{eqnarray} 
 H_{AT}(\{\sigma_i,\tau_i\}) &=&-\sum_{<ij>}  \left(  (K + K_4 \tau_i \tau_j )
\sigma_i \sigma_j  
 + K \tau_i \tau_j \right) \nonumber \\
 &=&-\sum_{<ij>}\left(  K_{ij}(\tau) \sigma_i \sigma_j  + K \tau_i \tau_j \right)
\nonumber \\
 &=&-\sum_{<ij>}\left(  \tilde{K}_{ij}(\sigma) \tau_i \tau_j  + K \sigma_i \sigma_j
\right) \; .
 \label{HAT2}
 \end{eqnarray} 
 one can build an ordinary FK cluster for the Ising model defined by
 the $\sigma$ (or $\tau$) spins which interact via the local couplings
 $K_{ij}(\tau)$ (or $\tilde{K}_{ij}(\sigma)$). This is just the basis of
 the cluster algorithm for the AT model employed in numerical
 simulations \cite{WD}.
 
 \begin{figure}[h]
 \includegraphics[scale=1.0,angle=0]{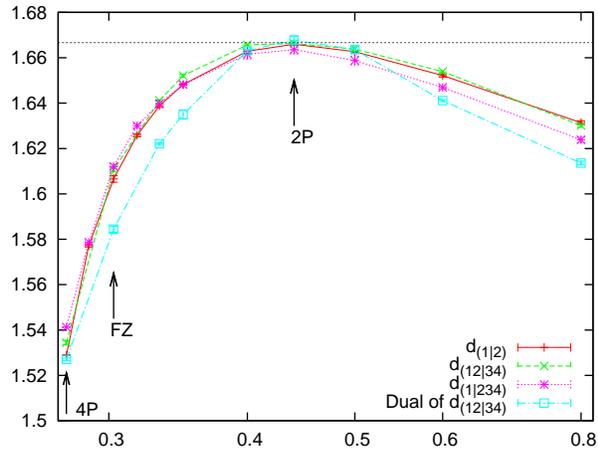}
 \caption{Fractal dimensions for the FK interfaces (defined in the
   text). We also show the dual of the fractal dimension for the spin
   interface with $(12|34)$ boundary condition.}  
 \label{PlotFKAT}  
 \end{figure}
 
 In order to build an FK interfaces, we proceeded as follows. For sake of
 clarity, let us consider the case with spin boundary conditions
 $(1|234)$. In Fig.~\ref{FKvsoth},
 a snapshot of a configuration with $(1|234)$ boundary conditions together 
  with the correspondent FK is shown. These conditions correspond to having
$\sigma_i=+$ and
 $\tau_i=+$ on half of the border. In this case it is equivalent to
 consider the Ising model in the $\sigma$ or in $\tau$ variables. Once
 we have chosen one of the two Ising model, we impose wired boundary
 conditions on one half of the border, i.e. the equal spins sitting on
 this half are linked by an FK bond with probability one.  After
 imposing these boundary conditions, we can build the FK cluster
 connected to the border and study its boundary. The fractal dimensions
$d^{FK}_{(1|234)}$
  of this  FK cluster is  shown in Fig.~\ref{PlotFKAT}. The results of  similar
construction for the
 boundary conditions $(1|2)$ and $(12|34)$ is also shown in Fig.~\ref{PlotFKAT}. 
 For the $(1|2)$ it is possible to build two types of
 interface, since the boundary conditions breaks the symmetry under the
 exchange of the two Ising models. We have checked numerically that the
 two types of interface produce the same fractal dimension.
 
 The results shown in Fig.~\ref{PlotFKAT} strongly support the fact
 that the values of the fractal dimension of the FK cluster are dual,
 see Eq.~(\ref{Duality}), to the fractal dimension $d_{(12|34)}$. At the
 FZ point the value of $d_{(1|234)}^{FK}$ is in strong agreement with
 the value $8/5$ which have been proposed on the basis of a CFT
 computation \cite{R}. At the Ising decoupling point this can be
 completely understood. Indeed remember that, at this point, the
 interface $d_{(12|34)}$ corresponds to the well known SLE$_{3}$
 interface of the Ising model which is also the spin cluster  boundary associated to 
 the FK cluster.  The  boundary of this
 cluster is described by an SLE$_{16/3}$ process and its fractal
 dimension is related to the spin interfaces one by the duality Eq.(\ref{Duality}).  In
 general,  the spin cluster  boundary associated to  all the FK cluster considered
here 
 in the interface whose fractal dimension is given by
 $d_{(12|34)}$. For the $(1|234)$ boundary condition for instance, 
 this can be clearly seen in Fig.~\ref{FKvsoth} where the FK cluster 
 and its outer boundary are shown. Note in particular that 
 the interface of the spin cluster on which th FK cluster is built and the spin
interface $(1|234)$ 
 are not the same.  
   
  Our numerical findings show then that the fractal
 dimension of the FK ans spin cluster boundary 
 are dual all along the critical line.  This is remarkable as it
 clearly shows that the Duplantier duality, which seems to be valid
 also for extended CFTs \cite{Pottsdis}, has a deep geometrical origin.
 
 Summing up, we have presented numerical results on spin cluster and FK
 cluster interfaces of the AT model which are generated by imposing
 certain boundary conditions. The AT model is particularly interesting
 in this respect as it presents a critical line in which the $Z_4$
 symmetry, together with the conformal one, plays a central role. We
 have computed the fractal dimensions of different spin interfaces all
 along the AT critical line.  The results are shown in
 Fig.~\ref{Plot2AT} : the main observation is that, for a general point
 on the critical line, the fractal dimension of a generic spin cluster
 interface takes one of four different possible values. These results
 interpolate between different interesting special critical models
 which can be found on the critical line, namely the $4$-states Potts,
 the FZ and the two-decoupled Ising point. At the FZ point the
 different values for the fractal dimensions were associated to the
 classification of conformal boundary conditions \cite{PS2}.  Another
 important finding is the existence of critical interfaces whose
 fractal dimension take the value $3/2$ all along the critical line:
 this result suggests the existence of interfaces described by $SLE_4$
 processes in the AT model, as it has been recently discussed in
 \cite{YM}.  Finally, we have computed the fractal dimension of the
 boundary of certain FK clusters, and the results are shown in
 Fig.~\ref{PlotFKAT}. All along the AT critical line, this value has
 been found to be the dual, see Eq.~(\ref{Duality}), to the fractal
 dimension of the boundary of the associated spin cluster, thus showing that
 Duplantier duality exists also for extended CFTs.
 
 The main motivation behind this work was to provide new insights into
 the geometrical properties of extended CFTs. We believe our results
 shed some light on the general problem of finding a systematic
 description of two dimensional critical phases in terms of stochastic
 geometry concepts.

 {\it Acknowledgements}:
 We acknowledge for the useful discussion and very helpful comments Y.~Ikhlef and
M. A. Rajabpour. We thank  Y.~Ikhlef
  for pointing us the results of \cite{AO} on the classification of the $(13|24)$
boundary conditions.

 \end{document}